# Polarizabilities and Effective Parameters for Collections of Spherical Nano-Particles Formed by Pairs of Concentric Double-Negative (DNG), Single-Negative (SNG) and/or Double-Positive (DPS) Metamaterial Layers

*Andrea Alù*[(1,2)], *and Nader Engheta*[(1),*]

[(1)] Department of Electrical and Systems Engineering, University of Pennsylvania

Philadelphia, PA 19104, U.S.A.

andreaal@ee.upenn.edu, engheta@ee.upenn.edu

[(2)] Department of Applied Electronics, University of Roma Tre

Via della Vasca Navale, 84 - 00146 Rome, Italy

alu@uniroma3.it

***ABSTRACT***

Unusual scattering effects from tiny spherical particles may be obtained when concentric shells are designed by pairing together "complementary" double-negative (DNG), single-negative (SNG), and/or standard double-positive (DPS) materials. By embedding these highly polarizable scatterers in a host medium one can achieve a bulk medium with interesting effective parameters. Some physical insights and justifications for the anomalous polarizability of these concentric spherical nano-particles and the effective parameters of the bulk composite medium are discussed.

* To whom correspondence should be addressed.

## INTRODUCTION

Double Negative (DNG) materials [1], also known as “left-handed” or “negative-index” media [2], are receiving increasing attention in the scientific community in recent years, due to the exciting properties of the wave interaction with these materials [1]-[21]. These features are consequences of the negative real part of their permittivity and permeability, which may be obtained by properly selecting and embedding some resonant inclusions in a host medium [3].

In our previous works, we have shown how ultracompact waveguides and resonators [4]-[6], anomalous wave tunneling and transparency [7]-[8] or dramatic enhancement of scattering from tiny spheres and cylinders [8]-[10] may be achieved when these materials are paired with conventional media (which, by analogy, can be referred to as “double positive (DPS)” materials). We have also shown that similar effects may be obtained using materials with *just* one of the two parameters being negative. These include ε-negative (ENG) media (such as plasmonic media) or the μ-negative (MNG) media, which possess, respectively, permittivity and permeability with a negative real part [5]-[10]. Again, when these two “conjugate” materials are judiciously paired together, the wave interaction shows anomalous properties. In those works, in particular, we have discussed how these anomalies may be explained as the effects due to the *interface resonance* taking place at the junction of materials with oppositely-signed constitutive parameters [7].

In one of our previous works [9], moreover, we have mentioned that when two concentric spherical nanoshells (or coaxial cylindrical nanoshells) made of DPS-DNG materials, or ENG-MNG, or even ENG-DPS or MNG-DPS materials, are suitably paired together, the overall scattering cross section of this tiny structure may be dramatically enhanced, as compared with the scattering from an object with the similar shape and size but made of standard DPS media only. In other words, one can obtain electrically small scatterers with a scattering cross section comparable with that of much larger scatterers. This implies that we can effectively have a “compact resonator”, i.e., an electrically small object in resonance. This phenomenon, again, may be explained using the concept of interface resonance when these conjugate materials are paired together.

In the present work, first we briefly discuss the general analysis of electromagnetic wave scattering from such spherical nanoparticles and their unusual scattering properties, highlighting the relationship between these scattering phenomena and the material polaritons, i.e., the natural modes supported by these particular scatterers. Other research groups have presented experimental results demonstrating these anomalous resonances in individual particles containing plasmonic materials, and have discussed certain theoretical aspects (e.g., hybridization) of this phenomenon [22]-[28]. Here, we give some other (different) features of the theory behind such scattering, and we then suggest the idea of embedding these tiny resonant particles with high polarizability as inclusions in a host material, in order to conceptually build a bulk composite medium, which may exhibit interesting characteristics with potential applications. The "compact resonant" inclusions, in fact, offer very high polarizabilities without occupying a large footprint, and consequently they may be easily embedded in a matrix array to produce an effective *hybrid-metamaterial* (i.e., a metamaterial whose inclusions may themselves be made of composite materials) with anomalous properties.

## SCATTERING PROPERTIES OF INDIVIDUAL METAMATERIAL NANOINCLUSIONS

The geometry of the single inclusion under study here is depicted in Fig. 1. This is a spherical scatterer composed of two concentric layers of radii $a_1$, $a$, with $a > a_1$, surrounded by a host material (with permittivity $\varepsilon_0$ and permeability $\mu_0$), which may not necessarily be free space. Let this scatterer be illuminated by an $e^{-i\omega t}$ – monochromatic plane wave. The concentric layers are assumed to be made of isotropic homogenous materials, which at the operating frequency $f = \omega / 2\pi$ have permittivities $\varepsilon_1$, $\varepsilon_2$ and permeabilities $\mu_1$, $\mu_2$, as indicated in the figure. In general, the constitutive parameters of the three material regions should be complex quantities, taking into account the possible material losses. Moreover, their real parts may be positive or negative, giving rise to the possibility of having DPS, DNG, ENG or MNG materials, as discussed in the introduction. Since the media are supposed to be passive, their imaginary parts are non-negative. Furthermore, in the following we assume lossless or low-loss conditions, i.e., $\mathrm{Im}\,\varepsilon_j \ll |\mathrm{Re}\,\varepsilon_j|$ and $\mathrm{Im}\,\mu_j \ll |\mathrm{Re}\,\mu_j|$, ($j = 0, 1, 2$).

Strictly speaking, every material except the free space exhibits temporal dispersion, and particularly metamaterials in the range of frequencies in which they display negative parameters may be strongly dispersive [1]-[3], [29]. Therefore, in the following discussion for the sake of simplicity we consider a monochromatic excitation, which allows the analysis to be performed for given values of the constitutive parameters at this specific frequency. The results for a single frequency may be promptly, but approximately, extended to a quasi-monochromatic case, typical of the telecommunication applications. In the next section, moreover, we will consider certain dispersion model for some metamaterials and plasmonic media.

Let an illuminating plane wave be propagating along the $\hat{\mathbf{z}}$ axis of the Cartesian coordinate system $(x, y, z)$ with its electric field directed along the $\hat{\mathbf{x}}$ axis. The scatterer is centered at the origin of this Cartesian and of a spherical coordinate system $(r, \theta, \varphi)$. As it is usually done, this plane wave can be written in terms of electric and magnetic vector potentials that are then expanded into spherical harmonics in the spherical coordinate system [30]-[31]:

$$\begin{aligned} \mathbf{E}_{inc} &= E_0 e^{ik_0 z}\hat{\mathbf{x}} = -\frac{E_0}{\varepsilon_0}\nabla\times\mathbf{F}_{inc} + \frac{iE_0}{\omega\mu_0\varepsilon_0}\nabla\times\nabla\times\mathbf{A}_{inc} \\ \mathbf{H}_{inc} &= \frac{E_0}{\eta_0} e^{ik_0 z}\hat{\mathbf{y}} = \frac{E_0}{\mu_0}\nabla\times\mathbf{A}_{inc} + \frac{iE_0}{\omega\mu_0\varepsilon_0}\nabla\times\nabla\times\mathbf{F}_{inc} \end{aligned}, \tag{1}$$

with:

$$\begin{aligned} \mathbf{A}_{inc} &= \hat{\mathbf{r}}\frac{\cos\phi}{\omega}\sum_{n=1}^{\infty} i^n \frac{2n+1}{n(n+1)} k_0\, r\, j_n(k_0 r) P_n^1(\cos\theta) \\ \mathbf{F}_{inc} &= \hat{\mathbf{r}}\frac{\sin\phi}{\omega\eta_0}\sum_{n=1}^{\infty} i^n \frac{2n+1}{n(n+1)} k_0\, r\, j_n(k_0 r) P_n^1(\cos\theta) \end{aligned}, \tag{2}$$

where $E_0$ is its complex amplitude, $j_n(.)$ are the spherical Bessel functions and $P_n^1(.)$ are associated Legendre polynomials of first degree and order *n* [32], $k_0 = \omega\sqrt{\mu_0\varepsilon_0}$ and $\eta_0 = \sqrt{\mu_0/\varepsilon_0}$ are the wave number and the characteristic impedance of the host medium, respectively. The correct choice for the signs of these square roots in the different types of media, depending on the sign of the real part of permittivity and permeability, is given in Table 1. For each type of medium, this has been obtained by adding an infinitesimally small amount of loss into the medium constitutive parameters, and then choosing the correct branch cut for the square roots in order to satisfy the radiation condition, as was similarly done for the DNG medium in [1].

Obviously, due to the linearity of the problem one may separately analyze the contribution of every term in the two summations. In particular, the series for the magnetic vector potential $\mathbf{A}_{inc}$ is composed of spherical $TM^r$ waves and analogously the electric potential $\mathbf{F}_{inc}$ is represented by a summation of $TE^r$ waves. (The superscript "r" will be dropped heretoafter for simplicity.)

This scattering problem is analogous to the one solved by Aden and Kerker [33], even though in their paper they have assumed to deal with conventional materials only (which we call DPS media). Their solution, however, is also applicable to the cases under study here, provided that the correct choice for the signs of the quantities $k_j = \omega\sqrt{\varepsilon_j \mu_j}$, $\eta_j = \sqrt{\mu_j / \varepsilon_j}$ ($j = 0, 1, 2$) is made for all the three media following Table 1.

In this work we are mostly interested in the field scattered by the particle (since from that we can evaluate the polarizabilities of this inclusion, to be embedded in a host matrix to form a bulk medium), which is determined for each spherical wave by the corresponding vector potential:

$$\begin{aligned} \mathbf{A}_s(n) &= \hat{\mathbf{r}} \frac{\cos\phi}{\omega} i^n \frac{2n+1}{n(n+1)} c_n^{TM} k_0 r\, h_n^{(1)}(k_0 r) P_n^1(\cos\theta) \\ \mathbf{F}_s(n) &= \hat{\mathbf{r}} \frac{\sin\phi}{\omega\eta_0} i^n \frac{2n+1}{n(n+1)} c_n^{TE} k_0 r\, h_n^{(1)}(k_0 r) P_n^1(\cos\theta) \end{aligned}, \tag{3}$$

where $h_n^{(1)}(.) = j_n(.) + i\, y_n(.)$ is the spherical Hankel function of the first kind. (This corresponds to $z_n^{(3)}(.)$ in the notation of [33]), and $y_n(.)$ here is the spherical Neumann function [32].) The scattering coefficients $c_n^{TE}$ and $c_n^{TM}$ (corresponding to $a_n^s$ and $b_n^s$ in [33] notation), which can be obtained when one applies the boundary conditions, completely determine the scattered field. They are given by Eqs. (26) and (27) in [33] and thus are not repeated here. The final expressions for the total scattered field are therefore given by:

$$\begin{aligned} \mathbf{E}_s &= \sum_{n=1}^{\infty} -\frac{E_0}{\varepsilon_0} \nabla\times\mathbf{F}_s(n) + \frac{iE_0}{\omega\mu_0\varepsilon_0} \nabla\times\nabla\times\mathbf{A}_s(n) \\ \mathbf{H}_s &= \sum_{n=1}^{\infty} \frac{E_0}{\mu_0} \nabla\times\mathbf{A}_s(n) + \frac{iE_0}{\omega\mu_0\varepsilon_0} \nabla\times\nabla\times\mathbf{F}_s(n) \end{aligned}, \quad r > a . \tag{4}$$

From the knowledge of the scattering coefficients, one can also express the total scattering cross section of this scatterer as [34]:

$$Q_s = \frac{2\pi}{|k_0|^2}\sum_{n=1}^{\infty}(2n+1)\left(\left|c_n^{TE}\right|^2 + \left|c_n^{TM}\right|^2\right), \tag{5}$$

and its backscattering cross section as:

$$\sigma = \pi\left|\frac{1}{k_0}\sum_{n=1}^{\infty}(-1)^n(2n+1)\left(c_n^{TE} - c_n^{TM}\right)\right|^2 . \tag{6}$$

It is worth noting that the vector potentials $\mathbf{A}_s(n)$ and $\boldsymbol{F}_s(n)$ of the scattered fields given in (3) represent the potentials for the radiation field of an electric and a magnetic multipole of order $n$, respectively, and the expressions given in (4), therefore, are equivalent to a multipole expansion [35]. This issue will be revisited later in the manuscript.

For the scattering coefficients $c_n^{TE}$ and $c_n^{TM}$, an equivalent, but physically more revealing expression than the one given in [33], may be obtained by manipulating the equations written for the boundary conditions at $r = a_1$ and $r = a$. They can be written, consistent with [36], as:

$$c_n^{TE} = -\frac{U_n^{TE}}{U_n^{TE} + iV_n^{TE}}, \quad c_n^{TM} = -\frac{U_n^{TM}}{U_n^{TM} + iV_n^{TM}}, \tag{7}$$

where the functions $U$ and $V$ are real valued for the lossless materials and for the TM-polarized scattered field they are given by the relations:

$$U_n^{TM} = \begin{vmatrix} j_n(k_1a_1) & j_n(k_2a_1) & y_n(k_2a_1) & 0 \\ \left[k_1a_1 j_n(k_1a_1)\right]'/\varepsilon_1 & \left[k_2a_1 j_n(k_2a_1)\right]'/\varepsilon_2 & \left[k_2a_1 y_n(k_2a_1)\right]'/\varepsilon_2 & 0 \\ 0 & j_n(k_2a) & y_n(k_2a) & j_n(k_0a) \\ 0 & \left[k_2a\, j_n(k_2a)\right]'/\varepsilon_2 & \left[k_2a\, y_n(k_2a)\right]'/\varepsilon_2 & \left[k_0a\, j_n(k_0a)\right]'/\varepsilon_0 \end{vmatrix}, \tag{8}$$

$$V_n^{TM} = \begin{vmatrix} j_n(k_1a_1) & j_n(k_2a_1) & y_n(k_2a_1) & 0 \\ \left[k_1a_1 j_n(k_1a_1)\right]'/\varepsilon_1 & \left[k_2a_1 j_n(k_2a_1)\right]'/\varepsilon_2 & \left[k_2a_1 y_n(k_2a_1)\right]'/\varepsilon_2 & 0 \\ 0 & j_n(k_2a) & y_n(k_2a) & y_n(k_0a) \\ 0 & \left[k_2a\, j_n(k_2a)\right]'/\varepsilon_2 & \left[k_2a\, y_n(k_2a)\right]'/\varepsilon_2 & \left[k_0a\, y_n(k_0a)\right]'/\varepsilon_0 \end{vmatrix}. \tag{9}$$

Analogous expressions for the TE polarization may be obtained by substituting $\varepsilon$ with $\mu$ into (8) and (9).

Expressions (7) show that $|c_n| \le 1$ and their peaks, which correspond to the scattering resonances for the structure, occur when $V_n = 0$ for the lossless materials. As in the case of a homogeneous scatterer [34], these peaks are related to the presence of material polaritons (i.e., natural modes) on the surface of the scatterer, which are responsible of the scattering resonances. In fact, $V_n^{TE} = Disp_n^{TE} = 0$ and $V_n^{TM} = Disp_n^{TM} = 0$ are indeed the dispersion relations for the TE and TM material polaritons for this scatterer, which may be obtained using the alternative technique proposed in [36], but applied here to this two-layer problem. (For a brief discussion about this technique for finding the resonant material polaritons for this structure and their physical relation with the scattering peaks, see Appendix A.)

When conventional DPS materials are considered, it is well known that such resonances (and the corresponding high scattering) may be obtained only when the size of the scatterer becomes comparable to the wavelength in the media [34]. Of course, this is due to the fact that all the material polaritons are below cut off when the size of the scatterer is less than a given dimension (similarly to what happens in an electrically small cavity), and as a result a small DPS scatterer leads to a large negative value of $V_n$ (as confirmed by Eq. (11)), and thus a very low value for $c_n$. Moreover, since $\lim_{a \to 0} U_n \to 0^-$ (as Eq. (10) will confirm), in small DPS scatterers (at least the ones with $\varepsilon > \varepsilon_0$) $\arg c_n \simeq \pi / 2$, following (7). When the size of the scatterer is increased, on the other hand, the magnitude of $c_n$ increases up to the resonance, which arises when the polariton is supported and $V_n = Disp_n = 0$, at which one gets $c_n = -1$ (with a $\pi / 2$ phase shift typical of any resonance phenomenon).

If instead of the DPS materials for all the media involved in the problem we consider a combination of DNG, SNG, and/or DPS metamaterials for the two layers in the geometry of Fig. 1, may this lead to unusual scattering properties, e.g., huge scattering resonances for such electrically small scatterers? In our previous works, we have theoretically verified how pairing *conjugate* (i.e., complementary) materials (i.e., materials with oppositely signed constitutive parameters) may indeed lead to the possibility of significant reduction in the lateral dimension of guided modes in planar and cylindrical waveguides and cavities [4]-[9]. In particular, we have shown that their dispersion relations in several different waveguide configurations filled by two of such conjugate materials do not depend on the *sum* of the thickness of the two regions, as it is the case for the usual waveguides, but on the *ratio* of thicknesses, implying the possibility of having sub-wavelength structures supporting guided modes with lateral dimension below the diffraction limitation [4]-[9]. These phenomena would indeed correspond to resonant modes in ultracompact cavities/waveguides, and we have explained how these resonances are directly related to pairing of such conjugate materials, and how they arise at the interface between them. In the present problem, an analogous situation is present: from the above analysis it follows that we should simply look for the possibility of support of resonant material polaritons by the nanospherical scatterers formed by a pair of DNG, SNG and/or DPS layers, even when their dimensions are much smaller than those required for the spheres made of standard DPS materials. It is worth noting how in [11] a similar goal has been achieved with a different technique, namely, increasing the effective material parameters (i.e., having high permittivity and/or high permeability) of the small scatterer, and thus decreasing the effective wavelength inside the material in order to make it comparable with its physical dimensions (In the cavity analogy this approach would correspond to filling the cavity with a high-permittivity material in order to make it resonant at a lower frequency). Of course, this process may be limited by the possibility of such an increase in the material parameters, i.e., a too small scatterer would require very high constitutive parameters and a geometrical limitation may always be present. Moreover, constructing materials with very high permittivities and/or permeabilities, which are often lossy, may involve certain technological challenges. In our approach, however, we will show below that, at least in principle, this physical limit may be overcome by exploiting the "resonant" pairing of the DNG, SNG, and/or DPS metamaterials, as already done in the other situations mentioned above, and at the same time the values of the constitutive parameters (and correspondingly their imaginary parts) do not need to be necessarily high.

For this purpose, let us analyze what happens when the size of the scatterer becomes very small, compared to the wavelength in every one of the three regions, i.e., when $|k_2| a \ll 1$, $|k_0| a \ll 1$, $|k_1| a_1 \ll 1$. In this limit, the expressions in (8) and (9) may be reduced to the following:

$$U_n^{TM} \simeq \frac{\pi (k_0 a)^{2n-1} (k_1/k_0)^n}{4^n (2n+1)^3 (n-1/2)!^2 k_2/k_0} \begin{vmatrix} 1 & 1 & -1 & 0 \\ (n+1)/\varepsilon_1 & (n+1)/\varepsilon_2 & n/\varepsilon_2 & 0 \\ 0 & \gamma^{-1} & -\gamma^{2n} & \gamma^{n-1} \\ 0 & (n+1)\gamma^{-n}/\varepsilon_2 & n\gamma^{n+1}/\varepsilon_2 & (n+1)/\varepsilon_0 \end{vmatrix}, \quad (10)$$

$$V_n^{TM} \simeq \frac{(k_1/k_0)^n}{(k_0 a)^2 (2n+1)^2 k_2/k_0} \begin{vmatrix} 1 & 1 & -1 & 0 \\ (n+1)/\varepsilon_1 & (n+1)/\varepsilon_2 & n/\varepsilon_2 & 0 \\ 0 & \gamma^{-1} & -\gamma^{2n} & -\gamma^{n-1} \\ 0 & (n+1)\gamma^{-n}/\varepsilon_2 & n\gamma^{n+1}/\varepsilon_2 & n/\varepsilon_0 \end{vmatrix}, \quad (11)$$

where $\gamma \equiv a_1/a$ is a shorthand for the ratio of the two radii. These closed form expressions reveal very interesting properties for electrically small scatterers. First, as already mentioned before, for small scatterers the value of $U_n$ is a small quantity and tends to zero as $(k_0 a)^{2n-1}$. (The determinants in (10) and (11) does depend on $\gamma$, but not on $a$ or $a_1$ separately.) This is consistent with the fact that usually a small scatterer has a very low intensity scattered field. On the other hand, as already anticipated, $V_n$ increases as $(k_o a)^{-2}$ when $a$ is reduced, owing to the fact that small scatterers made of conventional materials are far from supporting surface polaritons. In fact, for scatterers made of DPS materials the denominator in (7) is dominated by $V_n$ and the usual approximate expression for $c_n$ becomes:

$$c_n \cong i (k_0 a)^{2n+1} f_n(\gamma), \quad (12)$$

where $f_n(\gamma)$ is a positive function of $\gamma$. This confirms the well-known fact that for tiny scatterers made of DPS materials the wave scattering is weak and it is dominated by the first-order term (corresponding to the radiation of an electric or a magnetic dipole depending on the polarization we are considering), that the total scattering cross section decreases with $(k_0 a)^6$ and finally that the phase of the scattering coefficient is around $\pi/2$, which, as shown later, provides an equivalent electric dipole radiating in phase with the impinging excitation. These of course are all consistent with the conventional results expected in this case [34]. However, we should bear in mind that these assumptions are valid only if the determinant in $V_n$ is sufficiently larger than $U_n$, which is the case when small DPS scatterers are employed. If instead some combinations of DNG, SNG, and/or DPS metamaterials are used, either (or both) of the determinants in (10) and (11) may become zero in the small-radii formulation, where Eqs. (10) and (11) are applicable. The conditions for which the determinant in $V_n^{TM}$ in (11) becomes zero, and therefore a material polariton may be resonant in an electrically small scatterer and thus its scattering cross section may become comparable with that of a big sphere, are derived to be:

$$\begin{aligned} \text{TE: } \gamma \equiv \frac{a_1}{a} \simeq \sqrt[2n+1]{\frac{\left[(n+1)\mu_0 + n\mu_2\right]\left[(n+1)\mu_2 + n\mu_1\right]}{n(n+1)(\mu_2-\mu_0)(\mu_2-\mu_1)}} \\ \text{TM: } \gamma \equiv \frac{a_1}{a} \simeq \sqrt[2n+1]{\frac{\left[(n+1)\varepsilon_0 + n\varepsilon_2\right]\left[(n+1)\varepsilon_2 + n\varepsilon_1\right]}{n(n+1)(\varepsilon_2-\varepsilon_0)(\varepsilon_2-\varepsilon_1)}} \end{aligned}, \qquad (13)$$

which again depend only on $\gamma$ and not on the outer dimension of the scatterer. In the lossless case, these dispersion relations may be satisfied by some combinations of constitutive parameters, leading to the possibility of presence of material polaritons *independent* of the *total* size of the scatterer, but dependent on the *ratio* of the two radii. Of course we should take into account the physical limit $0 \le \gamma = a_1/a \le 1$, which implies the use of a combination of DNG, SNG, and/or DPS metamaterials in order to fulfill one or both of the conditions given in (13).

Fig. 2 illustrates the regions of permissible values for $(\varepsilon_1, \varepsilon_2)$ with corresponding values for $a_1/a$, in order to satisfy the condition (13) for the TM-polarized scattered wave, in the limit of lossless materials. The regions indicated with "brick" symbols are the "forbidden" regions for $(\varepsilon_1, \varepsilon_2)$, which cannot satisfy condition (13). (For example, the first quadrant, where both $\varepsilon_1$ and $\varepsilon_2$ are positive and represent conventional DPS materials, does not admit material polaritons for these spherical scatterers in the small-radii case. This confirms that the technique proposed in [11] to synthesize metamaterials with resonant tiny inclusions, that is to embed spherical inclusions with materials with high permittivity and permeability in a host medium, may indeed exhibit a physical limitation for the size of the scatterers.) In the regions other than the "brick" regions, a contour plot for $a_1/a$ has been plotted. Lighter regions correspond to higher values of $a_1/a$ (white for $a_1/a=1$, black for $a_1/a=0$). The corresponding figure for the TE-polarized scattered wave (not shown here) may be obtained by substituting $\varepsilon$ with $\mu$.

We note that the TM and TE conditions in (13) depend only on the permittivities and the permeabilities of the materials, respectively. This is due to the fact that in the limit of small radii (i.e., electrically small scatterer) we are approximately dealing with a quasi "electrostatic" and a quasi "magnetostatic" problem, and the electric and magnetic effects are effectively disjoint.

It is worth reiterating the fact that this resonant phenomenon relies on the judicious pairing of *conjugate* materials with oppositely signed constitutive parameters, at whose interface a local compact resonance may arise, similar to what we have studied in the other setups [4]-[9]. As already presented in [12], this "quasi-static" resonance may be viewed, at least for the first-order dipolar term, as a compact L-C resonance between the small positive capacitance represented by a DPS core (shell) and the small negative capacitance (which at the given operating frequency is effectively equivalent with a large inductance) of an ENG or DNG shell (core). Similar resonances in the scattering from nanospheres in terms of the ratio of radii have been predicted in [22] and experimentally shown in [23]-[27].

From formula (10) we may derive another condition for the effective "*transparency*" of the scatterer for the particular scattered mode. In the small-radii approximation, when the determinant in (10) vanishes, the scattering coefficient $c_n$ becomes zero. The conditions for this situation, which again depend only on the ratio of the two radii, are expressed as:

$$\begin{aligned} \text{TE: } \gamma \equiv \frac{a_1}{a} &= \sqrt[2n+1]{\frac{(\mu_2-\mu_0)\left[(n+1)\mu_2+n\,\mu_1\right]}{(\mu_2-\mu_1)\left[(n+1)\mu_2+n\,\mu_0\right]}} \\ \text{TM: } \gamma \equiv \frac{a_1}{a} &= \sqrt[2n+1]{\frac{(\varepsilon_2-\varepsilon_0)\left[(n+1)\varepsilon_2+n\,\varepsilon_1\right]}{(\varepsilon_2-\varepsilon_1)\left[(n+1)\varepsilon_2+n\,\varepsilon_0\right]}} \end{aligned}, \qquad (14)$$

Fig. 3 shows the permissible and "forbidden" regions for $\varepsilon_1$ and $\varepsilon_2$ to achieve the transparency for the TM-polarized scattered wave, and the corresponding values for $a_1/a$ in the permissible region, in analogy with Fig. 2.

It is interesting to note that in this case it is possible to achieve the transparency condition also for DPS-DPS

spheres, but one of the layers should be of a metamaterial (or plasmonic material) with $\varepsilon < \varepsilon_0$ or $\mu < \mu_0$, depending on the polarization of the scattered mode to be suppressed. We have discussed certain aspects of this property in a recent symposium [13], and the detailed results will be reported in the near future.

It should be mentioned here that the transparency conditions (14) are valid in the small-radii approximation, and one may argue that the scattering from small spheres is already small, if not zero. However, we may drastically reduce further the overall scattering cross section of small spheres by putting to zero the scattering coefficient for $n = 1$, since according to (12) all the other higher-order scattering coefficients are usually much lower than the first one. So if the dominant scattering coefficient can be made zero, the total scattering cross section of this sphere can be significantly reduced. For larger spheres an analogous cancellation of any given multipolar scattering coefficient may be achieved by a suitable choice of the two materials' parameters in order to make the determinant (8) zero. This of course reduces the total scattering cross section, but does not necessarily imply that the sphere becomes "completely" transparent, since for large spheres the overall scattering cross section may have non-zero contributions from several other multipolar scattering orders (see formulas (5) and (6)). We have analyzed this problem in detail, and the results will be included in a future publication.

An interesting point about conditions (13) is that they admit solution for $a_1 / a = 0$ and $a_1 / a = 1$, which implies the cases of a homogenous single-layer sphere. This means that the high scattering may be achieved also with such homogeneous spheres composed only of a single metamaterial, when the following conditions for the material parameters are satisfied:

$$\text{TE: } \mu_2 = -\frac{n+1}{n}\mu_0, \ \text{TM: } \varepsilon_2 = -\frac{n+1}{n}\varepsilon_0 . \qquad (15)$$

This effect is the well-known plasmonic resonance for a homogeneous sphere, which occurs for the metallic spherical nanoparticles at the optical or infra-red regimes [34], [37], and it depends on the modal order *n*. In particular, one can easily derive from (15) the usual condition for the resonant polariton of a homogeneous small sphere for the dipolar ($c_1^{TM}$) term, which is $\varepsilon = -2\,\varepsilon_0$.

Another issue to note about the "resonant" conditions (13) is that the high scattering effect may be achievable for any scattered mode with $n > 1$. This implies that one can, in principle, select the material parameters of the two layers of small sphere such that the dominant term with resonant scattering would be for the quadrupolar term (i.e., $n = 2$), the octopolar term (i.e., $n = 3$), or any higher-order terms, instead of having it for the dipolar one (i.e., $n = 1$). In other words, for such a case, a very small two-layer metamaterial nanosphere may strongly scatter as a quadrupole or an octopole, while its dipolar scattering may be much weaker. We have studied this issue in more detail, and have presented our preliminary results in a recent symposium [10]. The details of our analysis and findings will be reported elsewhere. Some experimental results and possible applications exploiting these second order or higher orders TM scattering resonances have been reported by other groups in [37]-[39].

As shown by the analysis reviewed above, when the ratio of radii is chosen to satisfy the "resonant" conditions (13), which implies that $V_n = Disp_n = 0$, the scattering coefficient $c_n$ attains its maximum magnitude. However, as we deviate from this "resonant" ratio, the value of $V_n$ for small radii becomes much higher than $U_n$, and hence the magnitude of coefficient $c_n$ drastically decreases. This suggests a certain "ratio bandwidth", which is reduced as the outer radius $a$ gets smaller and/or when higher scattering mode $n$ is considered (as follows from the expressions (10), (11)).

Fig. 4 shows the behavior of the resonant peaks of the scattering coefficient $c_1^{TM}$ for the small spherical scatterer with two concentric layers of ENG-DPS metamaterials. In particular, Fig. 4a shows the variation of $c_1^{TM}$ as one considers different outer radii $a$. As can be seen, when considering smaller $a$ the resonant peak of this scattering coefficient moves towards the value predicted by the approximate formula (13), and the "ratio bandwidth" becomes narrower. At the resonant peak, however, the scattering coefficient has the same value (of unity), independent of the total dimension of this nanosphere, and it resembles the scattering coefficient from an electrically large resonant sphere. This is seen in Fig. 4b, where the resonant peaks of the $c_1^{TM}$ coefficient are compared, in a logarithmic scale, for a DPS-ENG and a DPS-DPS scatterer. In other words, this nanosphere with two-layers of metamaterials acts as a "compact resonator", occupying a very small volume but exhibiting large resonant scattering cross section. The total and the back scattering cross sections in this case are dominated by the resonant coefficient $c_n$, and they can be written as:

$$Q_s \simeq \frac{2\pi(2n+1)}{|k_0|^2}, \quad \sigma \simeq \frac{\pi(2n+1)^2}{|k_0|^2}. \tag{16}$$

It is important to point out that for the small-radii approximation (i.e., as long as the other scattering terms in (5) and (6) are negligible) these cross sections are effectively independent of the outer radius at the polariton resonance. This implies that under condition (13) the far-zone scattered fields of these small scatterers are effectively independent of the outer radius of these spheres, and the far fields resemble those of the large spheres.

However, the near field distributions are different for different outer radii. When the scatterer dimensions are reduced, the field intensity around the scatterer indeed becomes extremely large, as expected (since the field is described by $y_n(.)$ functions). This is consistent with the strong field observed around metallic nanoparticles at the plasmonic resonance [23]-[28], which may lead to certain applications. Again, this effect is also analogous with what happens in a resonant cavity. As it may be seen from Fig. 4, reducing the overall dimension of the scatterer (i.e., the outer radius) effectively leads to a higher quality (Q) factor of the equivalent resonant cavity, and increases the reactive fields stored in the resonant mode.

Fig. 5 illustrates the near-zone distribution of the total electric field (Fig. 5a) and total magnetic fields (Fig. 5b) for the scatterer of Fig. 4a with $a = \lambda_0 / 100$ and $a_1 / a$ satisfying the resonant condition (13) for $n = 1$, under a plane wave excitation (as in formula (1)), normalized to the incident electric field amplitude and under the assumption of lossless materials. The normalized field strength shows large peak values around the scatterer (over 1000 times the amplitude of the impinging electric field), and it is dominated by the presence of the TM material resonant polariton (for $n = 1$), which is characterized by an electromagnetic field 90° out of phase with respect to the excitation, typical of any resonant phenomenon. As viewed in the far zone, the scatterer resembles a strong radiating electric dipole at the origin, somewhat similar to what would happen with a much larger resonant sphere with a conventional dielectric sphere. However, the large resonant sphere with DPS materials would also possibly exhibit contributions from higher-order terms, whereas for the small scatterer under study here, at the resonant condition (13), the scattering pattern is much more dominated by the dipolar pattern.

Clearly the material loss may sensibly affect the scattering properties of this scatterer, particularly for the smaller scatterers whose field distribution is highly concentrated in the materials. Fig. 6 shows the behavior of the resonant peaks when certain material loss is included in the analysis. The loss expectedly degrades and lowers the resonant peak, particularly for sharper resonances (for smaller $a$, as was just mentioned, and/or higher $n$). However, with a proper choice of metamaterial layers, the scattering strength may still be sensibly higher than that of the sphere of the same size but with conventional DPS materials, even when the losses are considered here. In the results shown in Fig. 6, the inner core is assumed to be lossless dielectric with $\varepsilon_1 = 1.2\varepsilon_0$ and the outer shell is silver with $\varepsilon_{Ag} = (-3.472 + i0.1864)\varepsilon_0$ at $\lambda_0 = 0.38\mu m$ [40]. To see the effect of material loss, the plots are shown for different values of imaginary part of permittivity of the outer shell, starting from the lossless case ($\varepsilon_i = 0$) up to the realistic value of this imaginary part for silver at $\lambda_0 = 0.38\mu m$ ($\varepsilon_i = 0.1864\varepsilon_0$). For comparison, the cases in which the outer shell is empty space (i.e., $\varepsilon_2 = \varepsilon_0$), in which it is made of the same lossless dielectric, and in which the entire particle is composed of silver are also shown. We can see that even with the realistic material loss included, the scattering is much stronger for the sphere with a pair of DPS-ENG layers, when compared with the DPS-DPS or ENG-ENG cases, due to the resonance phenomenon.

Finally, Fig. 7 shows the behavior of the resonant peaks for different scattering modes $n$ for the ENG-DPS sphere of Fig. 4a and the DPS-ENG sphere of Fig. 4b. As expected, the "bandwidth ratio" drastically decreases as one considers higher scattering mode $n$.

Analogous scattering effects may be expected for thin scatterers of other shapes formed by pairs of DNG, SNG, and/or DPS coaxial layers, as we have shown also in the cylindrical geometry [9]. A brief discussion of similar results for cylindrical nano-scatterers is given in Appendix B.

## POLARIZABILITIES AND EFFECTIVE PARAMETERS

Owing to their interesting scattering properties, a particulate composite medium formed by embedding many of these two-layered spherical nanoparticles as inclusions may exhibit unusual electromagnetic properties. If we limit ourselves to the resonances related to the dipolar scattering mode $n = 1$ for these small particles, the properties of the bulk medium can be described by its effective permittivity and permeability, which are related to the polarizability characteristics of the single particle and to the interaction constant of the whole lattice of particles (see e.g. [41]). In particular, we are interested in exploring the effects of the ratio of radii (and the excitation of the material polaritons) in such particles on the effective properties of the bulk medium.

We first need to express the polarizability tensors of the single spherical particle with two-layered metamaterials. As is well known, such tensors relate the local electromagnetic fields $(\boldsymbol{E}_{loc}, \boldsymbol{H}_{loc})$ to the induced dipole moments in a particle $(\boldsymbol{p}, \boldsymbol{m})$ [14], i.e.,

$$\begin{pmatrix} \mathbf{p} \\ \mathbf{m} \end{pmatrix} = \begin{pmatrix} \underline{\boldsymbol{\alpha}}_{ee} & \underline{\boldsymbol{\alpha}}_{eh} \\ \underline{\boldsymbol{\alpha}}_{he} & \underline{\boldsymbol{\alpha}}_{hh} \end{pmatrix} \cdot \begin{pmatrix} \mathbf{E}_{loc} \\ \mathbf{H}_{loc} \end{pmatrix} = \underline{\boldsymbol{\alpha}} \cdot \begin{pmatrix} \mathbf{E}_{loc} \\ \mathbf{H}_{loc} \end{pmatrix}. \tag{17}$$

Clearly the elements of the polarizability tensor $\underline{\boldsymbol{\alpha}}$ for the nanoparticles of interest in the present study are related to the scattering coefficients $c_1^{TM}$ and $c_1^{TE}$. After some mathematical steps, comparing the scattered fields in (4) for $n=1$ with the fields radiated by equivalent electric and magnetic dipoles placed at the origin, these relations can be expressed as:

$$\underline{\boldsymbol{\alpha}}_{ee} = -\frac{6\pi i \varepsilon_0 c_1^{TM}}{k_0^3} \underline{\mathbf{I}}, \quad \underline{\boldsymbol{\alpha}}_{he} = \underline{\mathbf{0}} \tag{18}$$

$$\underline{\boldsymbol{\alpha}}_{eh} = \underline{\mathbf{0}}, \quad \underline{\boldsymbol{\alpha}}_{hh} = -\frac{6\pi i c_1^{TE}}{k_0^3} \underline{\mathbf{I}}, \tag{19}$$

where the tensors $\underline{\mathbf{0}}$ and $\underline{\mathbf{I}}$ are the null and the identity tensors, respectively.

These expressions expectedly demonstrate that the TM scattering coefficient for $n=1$ is proportional to the electric dipole polarizability of the sphere and analogously the TE one is proportional to its magnetic dipole polarizability. They also satisfy the physical constraint [42]-[43] for particles with no material loss, i.e.,:

$$\operatorname{Im}\frac{1}{\alpha_{ee}} = -\frac{k_0^3}{6\pi\varepsilon_0}, \tag{20}$$

which from (18) implies that:

$$\operatorname{Re}\frac{1}{c_1^{TM}} = -1, \tag{21}$$

an expression consistent with (7).

In order to derive the effective constitutive parameters of a bulk medium composed of many of these identical particles as inclusions embedded in a host medium, we may utilize certain well-known mixing formulas [41]. If we assume that in the bulk medium the averaged distance between neighboring particles in any direction is small compared with the wavelength $\lambda_0$, and that the size of each scatterer is much smaller than this average distance, one can then apply the classical Clausius-Mosotti formula to estimate the effective parameters of the bulk the medium:

$$\varepsilon_{eff}^{r} = \varepsilon_0 + \frac{1}{\dfrac{1}{N\alpha_{ee}} - \dfrac{1}{3\varepsilon_0}}, \quad \varepsilon_{eff}^{p} = \varepsilon_0 + \frac{1}{\dfrac{1}{N}\left(\dfrac{1}{\alpha_{ee}} + \dfrac{i k_0^3}{6\pi\varepsilon_0}\right) - \dfrac{1}{3\varepsilon_0}}, \tag{22}$$

where $\varepsilon_{eff}^{r}$ is the effective permittivity for a random distribution of inclusions with number density $N$ (in which case the classical Maxwell-Garnett approach is applied [41] to obtain the first of (22)) and $\varepsilon_{eff}^{p}$ is the one for a periodic distribution with the same number density (in which case such an approach should be modified to cancel the radiation loss, as derived by Tretyakov and Viitanen [43], in order to obtain the valid second expression in (22)). Dual expressions for the effective permeability can be analogously derived (not shown here).

Let us examine the case where the inclusions are being operated as "compact resonators", i.e., when the TM condition (13) is satisfied for $n = 1$ (with no material loss). In this case, $c_1^{TM} = -1$, and thus $\alpha_{ee} = 6\pi i\varepsilon_0 k_0^{-3}$, which leads to the interesting relations:

$$\varepsilon_{eff}^{r} = \varepsilon_0 \frac{k_0^3 + 4i\pi N}{k_0^3 - 2i\pi N}, \quad \varepsilon_{eff}^{p} = -2\varepsilon_0 . \tag{23}$$

(As an aside, we note that at the resonance, the polarizability $\alpha_{ee} = 6\pi i\varepsilon_0 k_0^{-3}$, and consequently the expressions in (23), have no apparent information about the material parameters of the two metamaterial layers forming the particle. This is not surprising since for a given set of material parameters the ratio of radii is chosen to achieve condition (13), leading to $c_1^{TM} = -1$ -- an expression that does not depend on the distinct individual elements contributing to the resonance. This is analogous to what happens in any other resonant phenomenon (i.e., a resonant circuit), where, exactly at the resonance, the "outer" environment cannot infer the information about each of the individual single elements that are contributing to the resonance itself.)

From (20), Eq. (22) may be re-written for lossless particles as:

$$\varepsilon_{eff}^{r} = \varepsilon_0 + \frac{1}{\frac{1}{N}\mathrm{Re}\frac{1}{\alpha_{ee}} - \frac{1}{3\varepsilon_0}\left(1 + i\frac{k_0^3}{2\pi N}\right)}, \quad \varepsilon_{eff}^{p} = \varepsilon_0 + \frac{1}{\frac{1}{N}\mathrm{Re}\frac{1}{\alpha_{ee}} - \frac{1}{3\varepsilon_0}}. \tag{24}$$

This shows that $\varepsilon_{eff}^{r}$ for the case of random distribution approaches the value of $\varepsilon_{eff}^{p}$ for the case of periodic distribution when $N \gg k_0^3$ (an upper limit for $N$ however exists due to the range of validity of Eq. (22)). Consistent with the no-loss assumption, $\varepsilon_{eff}^{p}$ is a real quantity, whereas $\varepsilon_{eff}^{r}$ is indeed complex due to the scattering losses [41].

Concentrating on $\varepsilon_{eff}^{p}$, one notes that this medium may have a resonant permittivity when

$$\mathrm{Re}\frac{1}{\alpha_{ee}} = \frac{N}{3\varepsilon_0}, \text{ i.e., } \mathrm{Im}\frac{1}{c_1^{TM}} = -\frac{V_1^{TM}}{U_1^{TM}} = -\frac{2\pi N}{k_0^3}. \tag{25}$$

This resonance happens for a ratio $\left(\frac{a_1}{a}\right)_{\varepsilon}$ slightly different from the ratio $\left(\frac{a_1}{a}\right)$ that satisfies the particle resonant TM condition (13) for $n = 1$, but it gets closer to this number as $N$ becomes smaller and smaller. (Mathematically speaking, if we let $N \to 0$, we will have only a few particles, and thus the "medium" resonance will become similar to the "particle" resonance. Of course, as we make $N$ very small, we no longer have a "medium".) It should be noted that relations (24) are valid for not-extremely-dense concentrations of the inclusions in the host medium [41], which implies from (25) that the ratio $\frac{V_1^{TM}}{U_1^{TM}}$ of the single particle should not be very large at the resonance of the bulk medium. This of course implies that no bulk resonance can be expected in a medium composed of electrically small DPS inclusions, and that the resonance of the medium happens sufficiently close to the single-inclusion resonance, even though not exactly for the same ratio of radii. Needless to say, in order to satisfy the Kramers-Kronig conditions [29], the imaginary part of $\varepsilon_{eff}^{p}$ has a delta peak at this resonant value. Plots of $\varepsilon_{eff}^{p}$ and $\varepsilon_{eff}^{r}$ versus $a_1/a$ for the bulk medium formed by embedding particles with $\varepsilon_1 = 10\varepsilon_0$, $\varepsilon_2 = -1.5\varepsilon_0$ are shown in Fig. 8. In both cases, $a = \lambda_0/100$ and $N = (\lambda_0/10)^{-3}$. The behavior of the two plots is similar, except in the region near the resonance. The smaller $N$ gets, the less similar the two plots become. One notes that when the particle itself is at the resonance with the proper value for $a_1/a$, $\varepsilon_{eff}^{p}$ becomes $-2\varepsilon_0$, as mentioned before, and for this particular case this ratio is less than the value of $(a_1/a)_\varepsilon$ for which the bulk medium is at resonance. The effective permittivities shown in the figure tend to values close to $\varepsilon_0$ when $a_1/a$ is far away from the specific $a_1/a$ chosen for the particle resonance. This is expected because these inclusions are much smaller than the wavelength, and unless they are being operated at its resonance, i.e., act as "compact resonators" (or at the bulk medium resonance), their polarizabilities are very weak. In particular, we expect that $\varepsilon_{eff}^{p}(a_1/a = 0) \cong \varepsilon_{eff}^{r}(a_1/a = 0) = \varepsilon_0 + \Delta\varepsilon$, where $\Delta\varepsilon$ is a small positive number if $\varepsilon_2 > \varepsilon_0$ or $\varepsilon_2 < -2\varepsilon_0$, and a small negative number in other cases. This is due to the fact that the polarizability of a small homogenous sphere is given by:

$$\alpha \simeq 4\pi\varepsilon_0 a^3 \frac{\varepsilon - \varepsilon_0}{\varepsilon + 2\varepsilon_0}, \tag{26}$$

where $\varepsilon$ is the effective permittivity of the homogeneous sphere [41]. Therefore, when $\alpha$ is positive this should slightly increase the value of $\varepsilon_{eff}$ and when it is negative this should decrease it. Similar statements can be made about the limiting case of $a_1/a = 1$, but replacing $\varepsilon_2$ with $\varepsilon_1$ in the discussion.

We mentioned earlier that at the particle resonant ratio $a_1/a$ satisfying (13), the curve for $\varepsilon_{eff}^{p}$ passes through the point $\varepsilon_{eff}^{p} = -2\varepsilon_0$. Furthermore, if the material parameters of the two layers in the inclusion particle are in the permissible region of the plot in Fig. 3 for the transparency condition and if $a_1/a$ satisfies the TM portion of condition (14), the effective permittivity $\varepsilon_{eff}^{p}$ should clearly become equal to $\varepsilon_0$. This is expected since for this ratio of radii the electric polarizability of each particle is zero (i.e., the particles are effectively transparent), and thus the bulk medium will be transparent as well.

From Eqs. (13) and (14), one can also conclude that $\left(a_1/a\right)^{TM}_{resonance} > \left(a_1/a\right)^{TM}_{transparency}$ when $\varepsilon_2 > 0$, and $\left(a_1/a\right)^{TM}_{resonance} < \left(a_1/a\right)^{TM}_{transparency}$ when $\varepsilon_2 < 0$ (of course in the regions where the two ratios have physical meaning). Starting from these inequalities and the previous considerations, it is straightforward to predict heuristically the behavior of the monotonic curve in Fig. 8a, in terms of the various ranges of $\varepsilon_1$ and $\varepsilon_2$, as summarized in Table 2.

We may also examine the behavior of these plots in terms of the frequency of operation. For instance, if we take the outer layer of the spherical particle to be a standard dielectric material with $\varepsilon_2 = 10\varepsilon_0$ and the inner core to be a lossless plasmonic medium (i.e., an ENG medium) with the Drude model for its permittivity (neglecting losses), i.e., $\varepsilon_1 = \varepsilon_0 \left(1 - \frac{\omega_p^2}{\omega^2}\right)$, we can describe the effective permittivity in terms of frequency. Fig. 8 demonstrates such results for a given $a_1/a = 0.9$.

Obviously a similar analysis may be performed for the effective permeability of the bulk medium, which depends only on the permeabilities of the two metamaterials composing each nano-spherical inclusion (due to the quasi-static behavior of the field inside the nanosphere). The TE material polariton may be resonantly excited in these tiny particles and analogous relationships similar to the ones derived above may be obtained here, with the usual replacement of $\varepsilon$ with $\mu$ in the discussion. Overall, this might represent, at least conceptually, an interesting new venue for manufacturing isotropic metamaterials with negative permittivities and/or permeabilities. It is worth noting that by the same means we may synthesize bulk materials having low or zero $\varepsilon$ and/or $\mu$ (with low amount of losses), thus having near-zero refractive index. This may open up several possibilities for different applications, as shown for instance in [15]-[21].

**SUMMARY**

Unusual scattering effects from tiny spherical particles may be obtained when concentric shells are designed by pairing together "complementary" double-negative (DNG), single-negative (SNG), and/or standard double-positive (DPS) materials. By embedding these highly polarizable scatterers in a host medium one can achieve a bulk medium with interesting effective parameters. In this work, we first reviewed and discussed various scattering characteristics of small spherical particles made of pairs of DNG, SNG, and/or DPS metamaterials, and provided certain physical insights into the mathematical descriptions of these scattering phenomena. Since these particles may indeed act as "compact resonators", i.e., scatterers with strong resonant scattering cross section but a very small physical volume, they can be good candidates for inclusions in constructing particulate composite media. The effective permittivity and permeability of such bulk media have been discussed here and it has been shown how they depend on various parameters of the two-layered particle inclusions. The effective parameters of these bulk media can thus be adjusted and tailored using several degrees of freedom, thus providing possibilities for constructing more complex metamaterials. Since in the optical and IR domains, in particular, some noble metals, e.g., silver, aluminum and gold, exhibit negative values for the real part of their permittivities, such metals can be used as the dielectric-coated metallic (or metal-coated dielectric) spherical nanoparticles for use in constructing the bulk media discussed here, which may lead to metamaterials with negative bulk permittivity or permeability.

## APPENDIX A: MATERIAL POLARITON

Following the technique described in [36], one can find the material polariton for the spherical scatterer shown in Fig. 1. First we conceptually surround this scatterer with a spherical metal wall of very large radius, i.e., $r = R \gg a$, and then we look for those resonant modes of this "big cavity" with a field distribution mainly concentrated around the scatterer, which are described by the $y_n\left(k_0 r\right)$ functions for their radial dependence since the scatterer is small and placed at the origin. Since the determinant in (9) is exactly the one obtained when the eigensolutions for these modes are derived, $V_n = Disp_n = 0$ corresponds to the dispersion relation (for the TE or TM modes) of the material polaritons supported by this scatterer.

It is interesting to note that in the scattering problem, when the object is illuminated by an external wave, for the combination of parameters for which $Disp_n$ vanishes, the total field in the host medium resembles exactly the material polariton distribution we have just described in the “big cavity”, with a 90 degrees phase shift with respect to the excitation. Specifically, if the *n*-th spherical harmonic component of the impinging field is written as $\zeta\, j_n\left(k_0 r\right)$, with $\zeta$ a generic complex quantity, the scattered field becomes $\zeta\, c_n h_n^{(2)}\left(k_0 r\right) = -\zeta h_n^{(2)}\left(k_0 r\right) = -\zeta\, j_n\left(k_0 r\right) + j\,\zeta\, y_n\left(k_0 r\right)$ (since $c_n = -1$ when $V_n = Disp_n = 0$, following (7)). Summing the two expressions in order to derive the total field in the host medium, the field in the outer region shows *only* the material polariton distribution $j\,\zeta\, y_n\left(k_0 r\right)$, with the 90 degrees phase difference with respect to the incident field. This has been shown clearly in the field distributions of Fig. 5 and it is analogous to what happens in a L-C circuit when it is driven at the resonant frequency or when a surface wave mode is excited by an impinging evanescent wave.

By analogy, we may relate the expression for $U_n$ in (8) to those resonant modes of the big spherical cavity that are described only by $j_n\left(k_0 r\right)$ functions. In this case, $U_n = 0$ represent the dispersion relations for such modes (note how the expressions of $U_n$ and $V_n$ differ just for the terms $j_n\left(k_0 a\right)$ and $y_n\left(k_0 a\right)$ in (8) and (9)), which by analogy may be called vacuum polaritons [36]. It is not a coincidence, of course, that in the big cavity these modes are less affected by the presence of the scatterer, since their field goes to zero at the origin, and in the scattering problem the scattered field disappears when $U_n = 0$ (since the corresponding $c_n$ is also zero). This might be an interesting physical insight into the anomalous transparency effect we have discussed in a recent symposium [13], and will be reported in more detail in a future publication.

## APPENDIX B: CYLINDRICAL CASE

Similar to what we have shown here for the spherical geometry, also ellipsoidal or cylindrical inclusions may be treated in the same analytical way, and may conceptually yield similar results. In the cylindrical reference system, for instance, a 2-D problem (for normal incidence) similar to the one presented here may be solved, yielding to the excitation of resonant cylindrical polariton in nano-rods. The analytical treatment is not shown here, but it clearly involves cylindrical Bessel functions instead of the spherical Bessel functions used here. The formulas analogous to (13) in this case are given as [9]:

$$\begin{aligned} &\text{TE: } \gamma \equiv \frac{a_1}{a} \simeq \sqrt[2n]{\frac{\mu_2+\mu_1}{\mu_2-\mu_1}\frac{\mu_2+\mu_0}{\mu_2-\mu_0}} \\ &\text{TM: } \gamma \equiv \frac{a_1}{a} \simeq \sqrt[2n]{\frac{\varepsilon_2+\varepsilon_1}{\varepsilon_2-\varepsilon_1}\frac{\varepsilon_2+\varepsilon_0}{\varepsilon_2-\varepsilon_0}} \end{aligned} \quad . \tag{27}$$

Also in this case we may predict anomalous effects for a bulk medium consisting of collections of these resonant nano-rods and the results would predictably be analogous to the ones presented here.

**TABLES**

Table 1 – Choice of the sign for the square roots in the expressions of the characteristic impedances $\eta$ and of the wave numbers *k* in the different types of media.

| | | |
|---|---|---|
| DPS ( $\mathrm{Re}\,\varepsilon > 0$ , $\mathrm{Re}\,\mu > 0$ ) | $\eta > 0$ | $k > 0$ |
| DNG ( $\mathrm{Re}\,\varepsilon < 0$ , $\mathrm{Re}\,\mu < 0$ ) | $\eta > 0$ | $k < 0$ |
| ENG ( $\mathrm{Re}\,\varepsilon < 0$ , $\mathrm{Re}\,\mu > 0$ ) | $\eta \in \Im$ , $\mathrm{Im}\,\eta < 0$ | $k \in \Im$ , $\mathrm{Im}\,k > 0$ |
| MNG ( $\mathrm{Re}\,\varepsilon > 0$ , $\mathrm{Re}\,\mu < 0$ ) | $\eta \in \Im$ , $\mathrm{Im}\,\eta > 0$ | $k \in \Im$ , $\mathrm{Im}\,k > 0$ |

Table 2 – Behavior of the curve $\varepsilon_{eff}^{p}$ in terms of the various possible combinations of material permittivities for which the bulk medium shows a resonant behavior.

| **Admissible regions for the resonant inclusions (as in Fig. 2)** | $a_1/a=0$ | $a_1/a=1$ | $(a_1/a)^{TM}_{transp.}$ | $(a_1/a)_{resonance-medium}$ | $\partial\varepsilon_{eff}^{p}/\partial(a_1/a)$ |
|---|---|---|---|---|---|
| $\varepsilon_1<-2\varepsilon_0$ and $\varepsilon_2>-\varepsilon_1/2$ | $\varepsilon_{eff}^{p}>\varepsilon_0$ | $\varepsilon_{eff}^{p}>\varepsilon_0$ | $<(a_1/a)^{TM}_{resonance}$ | $>(a_1/a)^{TM}_{resonance}$ | $<0$ |
| $-2\varepsilon_0<\varepsilon_1<0$ and $0<\varepsilon_2<-\varepsilon_1/2$ | $\varepsilon_{eff}^{p}<\varepsilon_0$ | $\varepsilon_{eff}^{p}<\varepsilon_0$ | $<(a_1/a)^{TM}_{resonance}$ | $<(a_1/a)^{TM}_{resonance}$ | $>0$ |
| $-2\varepsilon_0<\varepsilon_2<0$ and $(\varepsilon_1<-2\varepsilon_0)\vee(\varepsilon_1>\max(\varepsilon_0,-2\varepsilon_2))$ | $\varepsilon_{eff}^{p}<\varepsilon_0$ | $\varepsilon_{eff}^{p}>\varepsilon_0$ | NO | $>(a_1/a)^{TM}_{resonance}$ | $<0$ |
| $-\varepsilon_0/2<\varepsilon_2<0$ and $-2\varepsilon_2<\varepsilon_1<\varepsilon_0$ | $\varepsilon_{eff}^{p}<\varepsilon_0$ | $\varepsilon_{eff}^{p}<\varepsilon_0$ | $>(a_1/a)^{TM}_{resonance}$ | $>(a_1/a)^{TM}_{resonance}$ | $<0$ |
| $\varepsilon_2<-2\varepsilon_0$ and $-2\varepsilon_0<\varepsilon_1<\varepsilon_0$ | $\varepsilon_{eff}^{p}>\varepsilon_0$ | $\varepsilon_{eff}^{p}<\varepsilon_0$ | NO | $<(a_1/a)^{TM}_{resonance}$ | $>0$ |
| $\varepsilon_2<-2\varepsilon_0$ and $\varepsilon_0<\varepsilon_1<-2\varepsilon_2$ | $\varepsilon_{eff}^{p}>\varepsilon_0$ | $\varepsilon_{eff}^{p}>\varepsilon_0$ | $>(a_1/a)^{TM}_{resonance}$ | $<(a_1/a)^{TM}_{resonance}$ | $>0$ |

**FIGURE CAPTIONS**

Figure 1 – Cross section of a spherical nano-particle (as inclusion) composed of two concentric layers of different isotropic materials in a suitable spherical reference system $(r, \theta, \varphi)$.

Figure 2 – Regions for which the dispersion relation (13) for the TM-polarized scattered wave is satisfied, with the corresponding values for $a_1/a$ between zero and unity. The "forbidden" regions indicted with the "brick" symbols present values of $\varepsilon_1$ and $\varepsilon_2$ for which the condition (13) for the TM case cannot be fulfilled.

Figure 3 – Regions for which the transparency condition (14) for the TM-polarized scattered wave is fulfilled, with the corresponding values for $a_1/a$ between zero and unity. Analogous to Fig. 2, the "forbidden" regions indicated with the "brick" symbols correspond to values of $\varepsilon_1$ and $\varepsilon_2$ for which the condition (14) for the TM case has no physical meaning.

Figure 4 –Magnitude of the scattering coefficient $c_1^{TM}$ versus the ratio of radii $a_1/a$: in (a), for the ENG-DPS

spherical scatterer with $\varepsilon_1 = -3\varepsilon_0$, $\varepsilon_2 = 10\varepsilon_0$, $\mu_1 = \mu_2 = \mu_0$ with the outer radius $a$ as a parameter; in (b), for $a_1 = 0.01\lambda_0$, $\varepsilon_1 = 10\varepsilon_0$, $\varepsilon_2 = \pm 1.2\varepsilon_0$, $\mu_1 = \mu_2 = \mu_0$, as a comparison with a DPS-DPS case (logarithmic scale).

Figure 5 – Normalized near-zone total electric field (a) and magnetic field (b) distribution for a two-layer spherical scatterer with $\varepsilon_1 = -3\varepsilon_0$, $\varepsilon_2 = 10\varepsilon_0$, $\mu_1 = \mu_2 = \mu_0$, $a = 0.01\lambda_0$ and $a_1$ chosen to satisfy condition (13) for the TM case $n = 1$, i.e., when $Disp_1^{TM} = 0$. They are shown both on the cut $\theta = \pi/2$, $\phi = \pi/2$, where only the $E_\theta$ and $H_\phi$ component are present, showing real and imaginary parts, and on the plane *x*-*y* with their absolute value. Owing to the resonant condition, the electric and magnetic field distributions are $\pi/2$ out of phase with respect to the incident electric and magnetic fields, respectively.

Figure 6 – Magnitude of the scattering coefficient $c_1^{TM}$, when material loss for the outer shell is included. In this case $\varepsilon_1 = 1.2\varepsilon_0$, $\varepsilon_2 = -3.472\varepsilon_0$, $\mu_1 = \mu_2 = \mu_0$, $a = \lambda_0/20$. The imaginary part of $\varepsilon_2$ is selected for each curve, and for the dot-line $\varepsilon_2$ corresponds to the real value of silver at the free space wavelength $\lambda_0 = 0.38\mu m$. In the DPS-DPS case $\varepsilon_1 = \varepsilon_2 = 1.2\varepsilon_0$, in the DPS-air case the outer shell is not present and in the "all Ag" case the sphere has $\varepsilon_1 = \varepsilon_2 = (-3.472 + i0.1864)\varepsilon_0$.

Figure 7 – Magnitude of the scattering coefficient $c_n^{TM}$ for different scattering modes $n$ for a sphere with $\varepsilon_1 = 10\varepsilon_0$, $\varepsilon_2 = -1.2\varepsilon_0$, $\mu_1 = \mu_2 = \mu_0$, $a_1 = 0.05\lambda_0$.

Figure 8 – Plot of a) $\varepsilon_{eff}^{p}$ and b) $\varepsilon_{eff}^{r}$ for a bulk medium constructed by embedding many identical spherical particles with $\varepsilon_1 = 10\varepsilon_0$, $\varepsilon_2 = -1.5\varepsilon_0$, $a = \lambda_0/100$ and $N = (\lambda_0/10)^{-3}$.

Figure 9: Plot of a) $\varepsilon_{eff}^{p}$ and b) $\varepsilon_{eff}^{r}$ for a bulk medium constructed by embedding many identical spherical particles with $\varepsilon_1 = \varepsilon_0\left(1 - \omega_p^2/\omega^2\right)$, $\varepsilon_2 = 10\varepsilon_0$, $a = \lambda_0/100$, $a_1/a = 0.9$ and $N = (\lambda_0/10)^{-3}$.

**FIGURES**

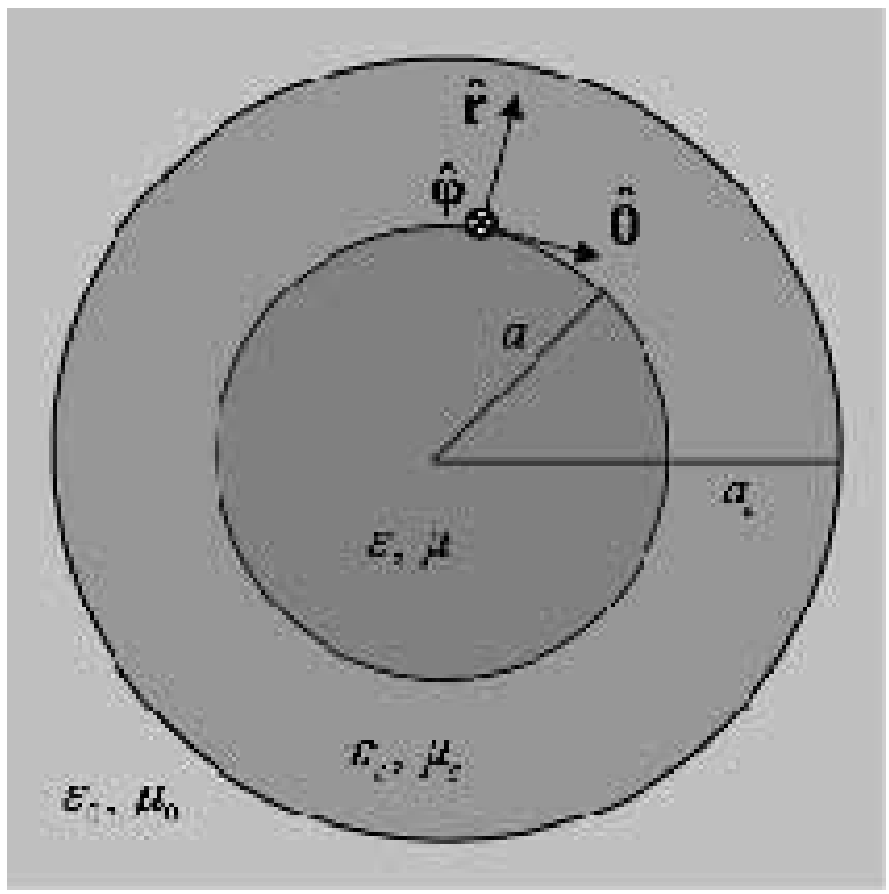

Figure 1

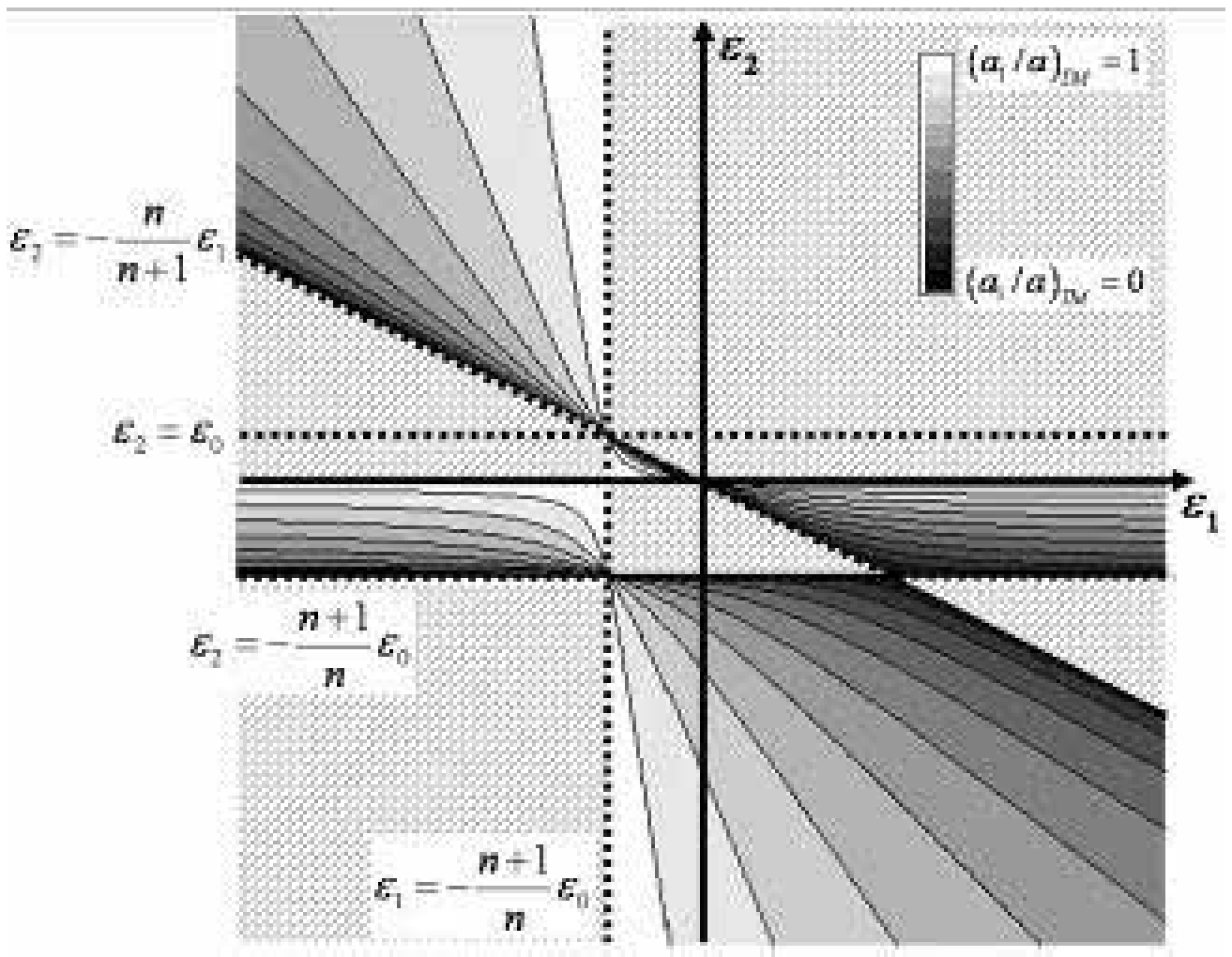

Figure 2

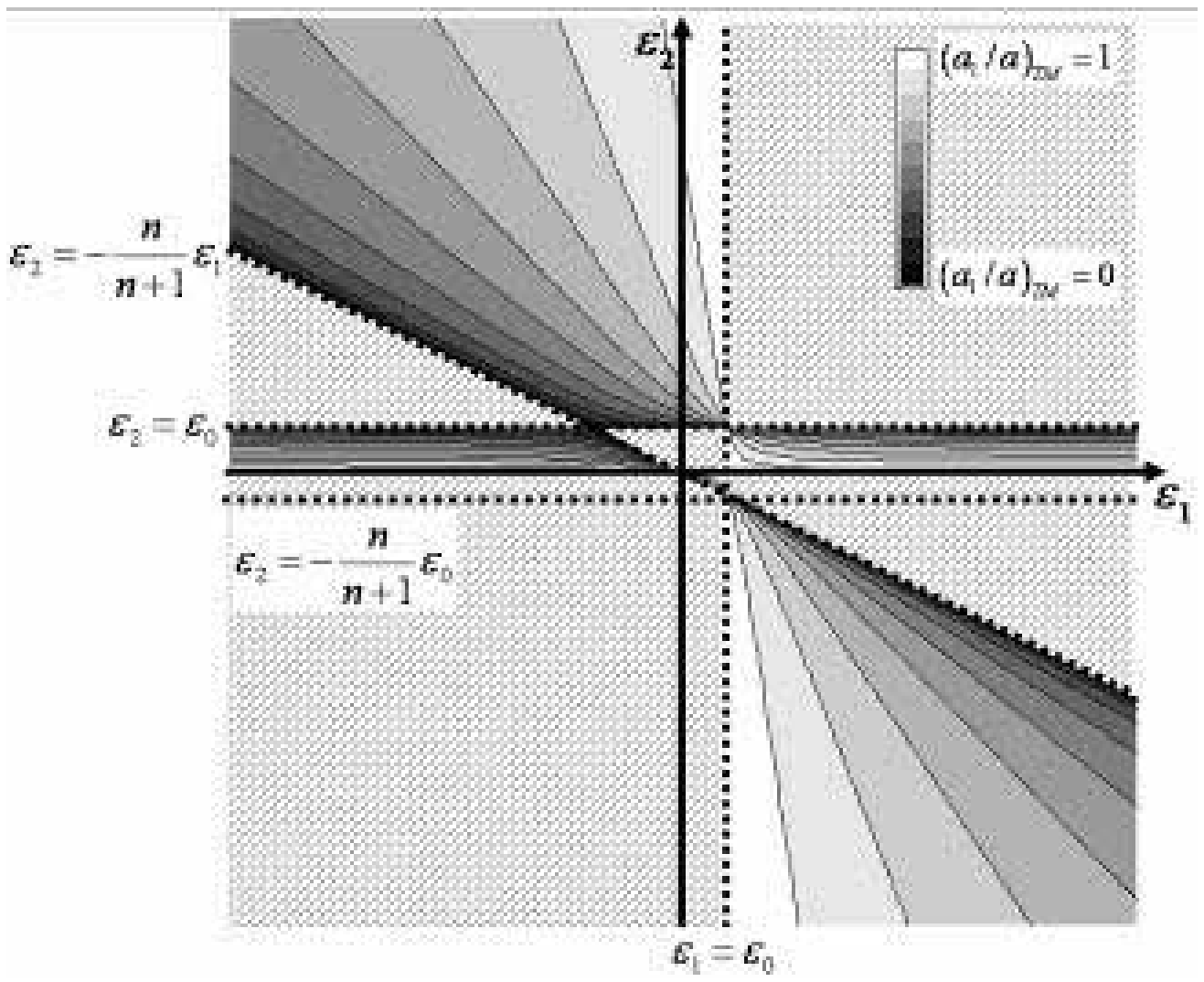

Figure 3

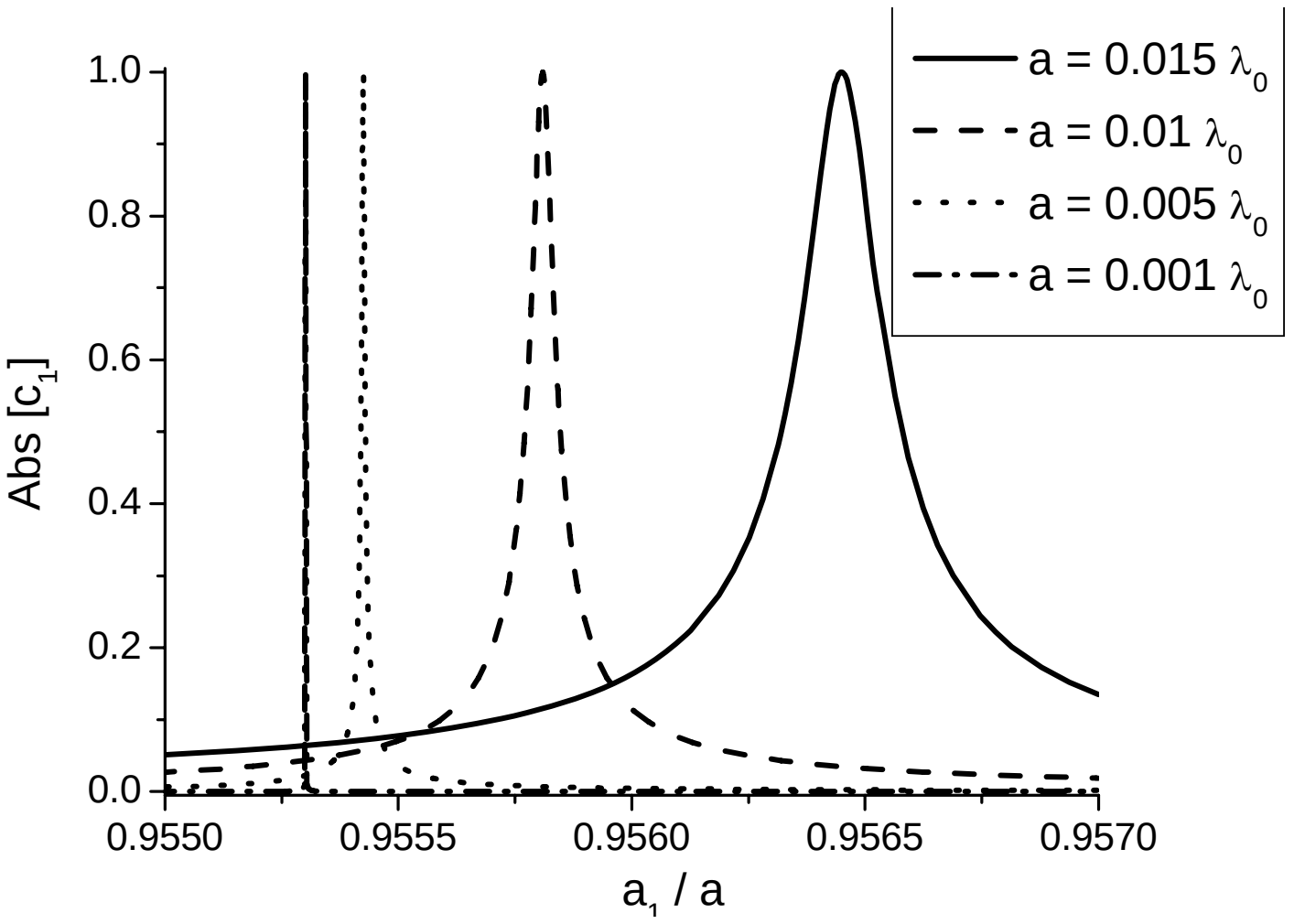

a = 0.015 λ0
a = 0.01 λ0
a = 0.005 λ0
a = 0.001 λ0
Abs [c1]
a1 / a

a)

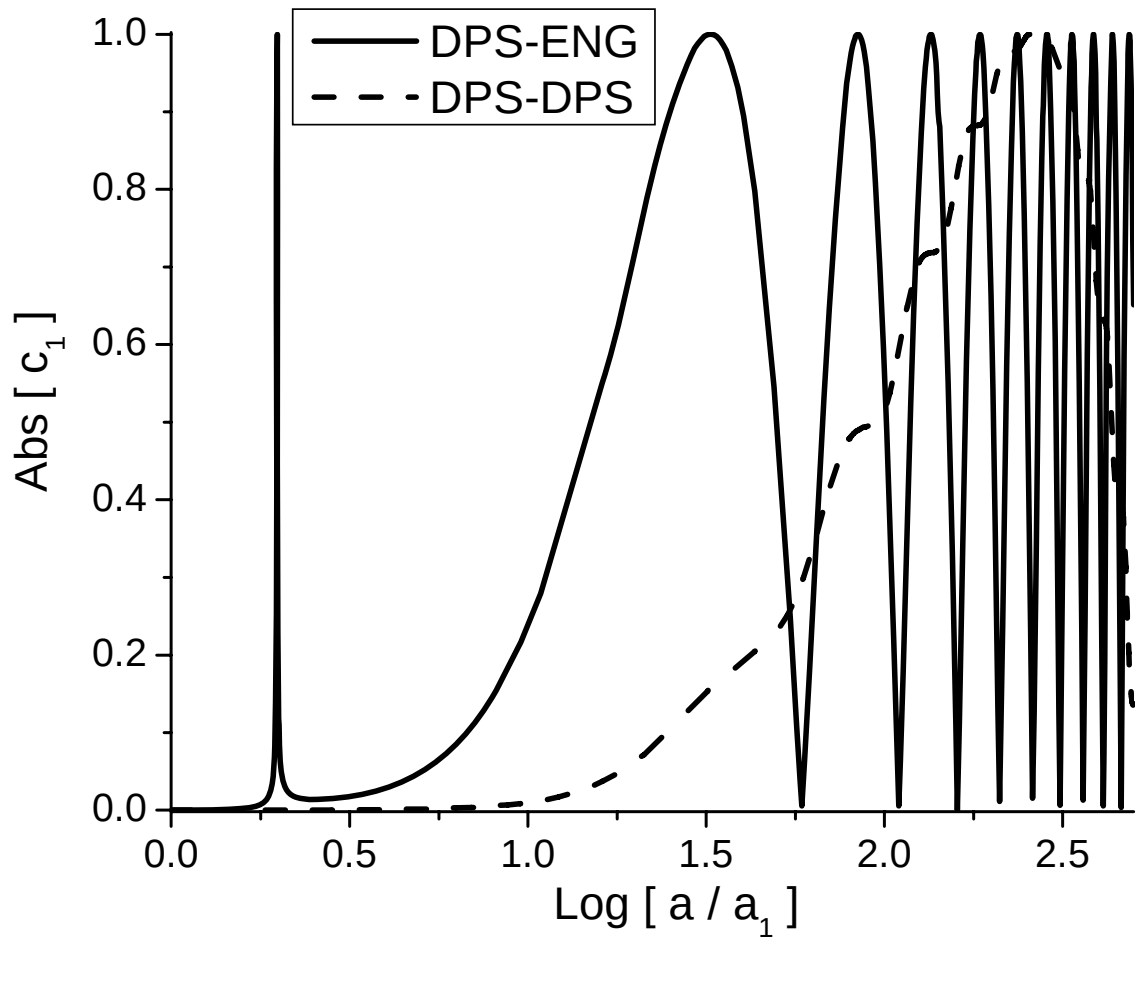

DPS-ENG
DPS-DPS
Abs [ c1 ]
Log [ a / a1 ]

b)

Figures 4a and 4b

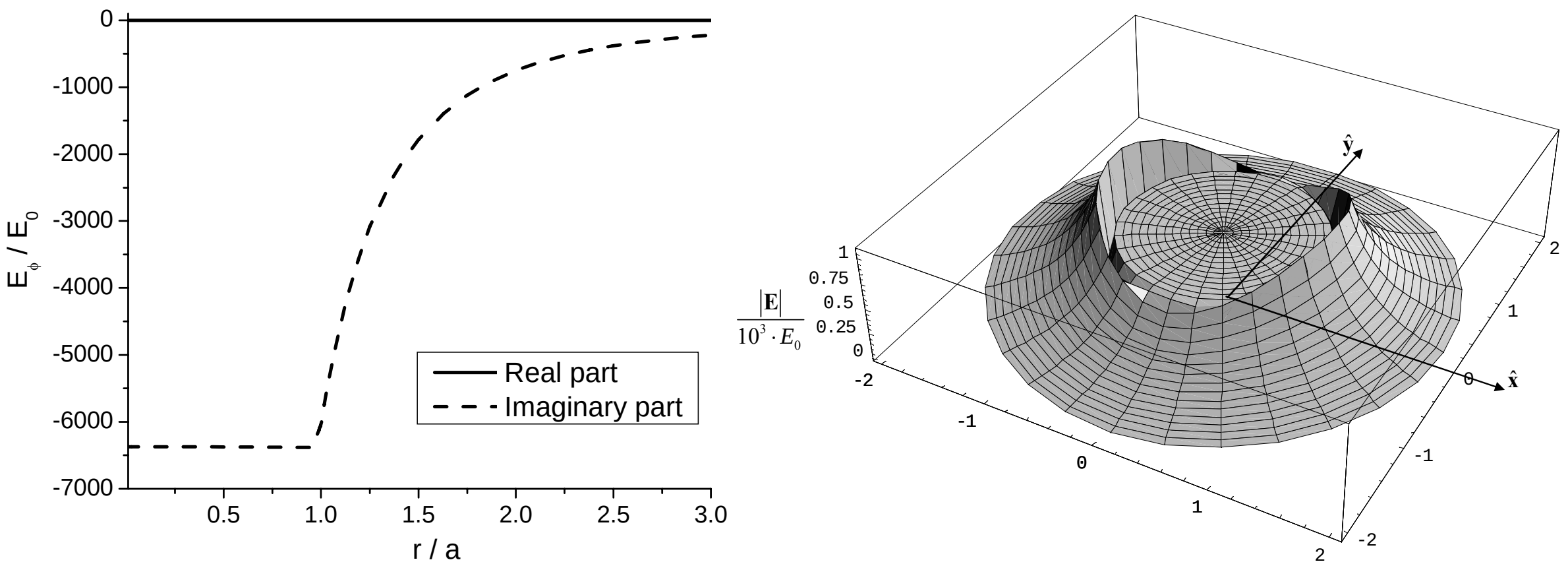

a)

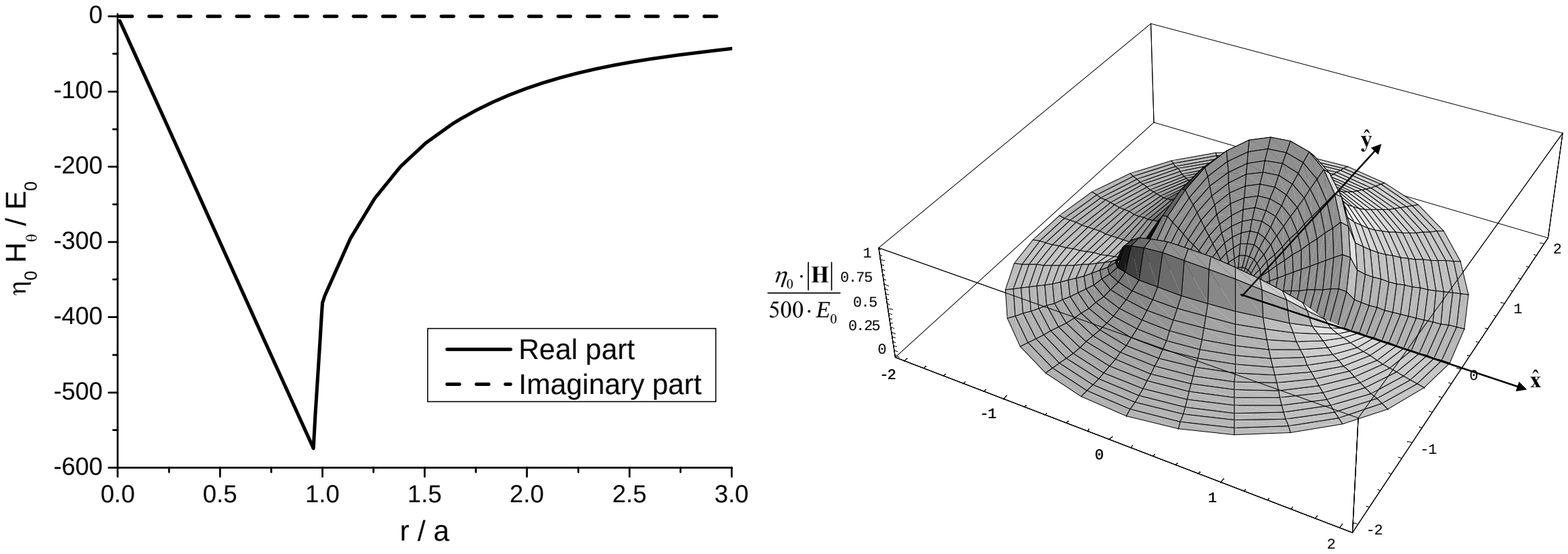

b)

Figures 5a and 5b

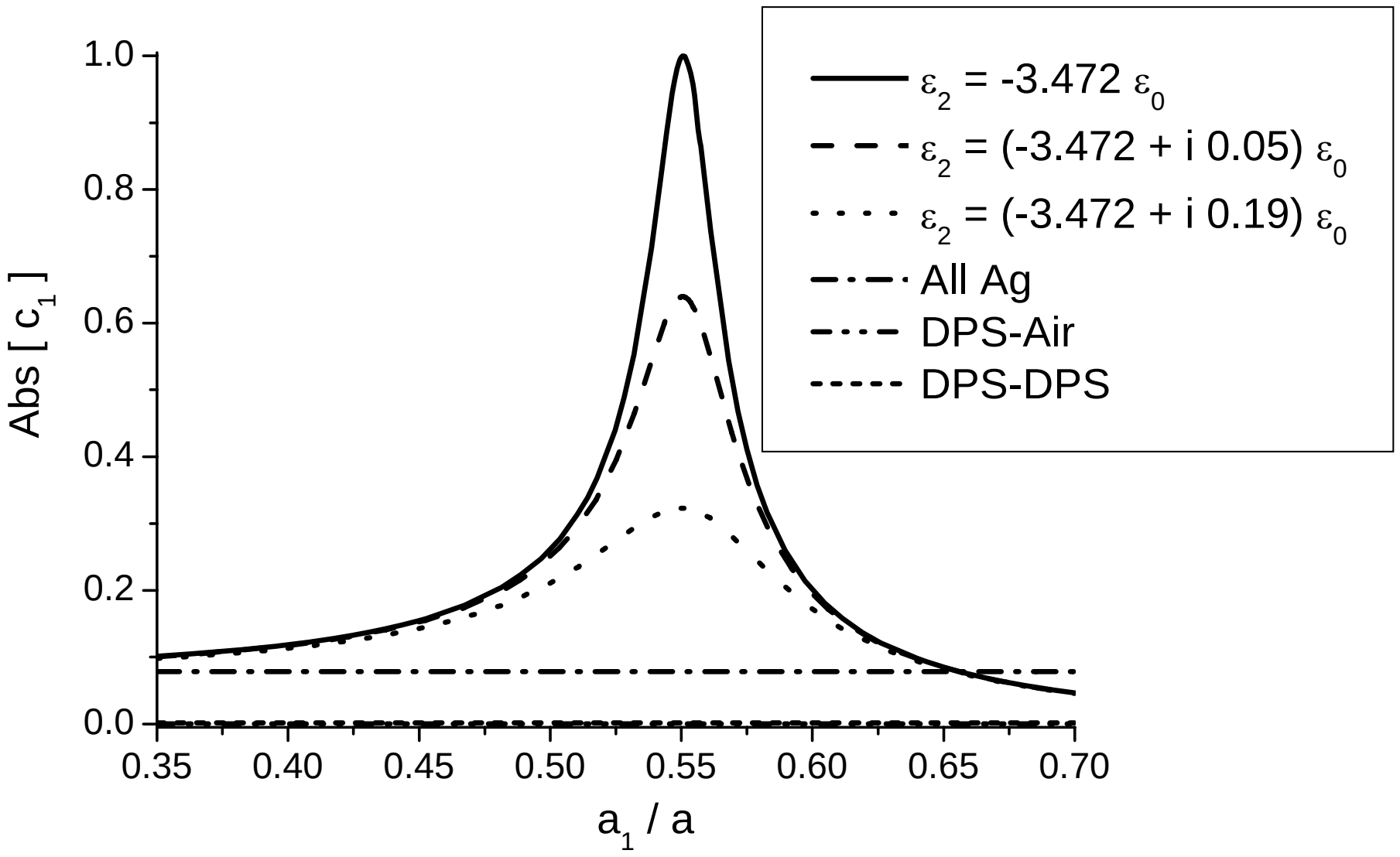

Figure 6

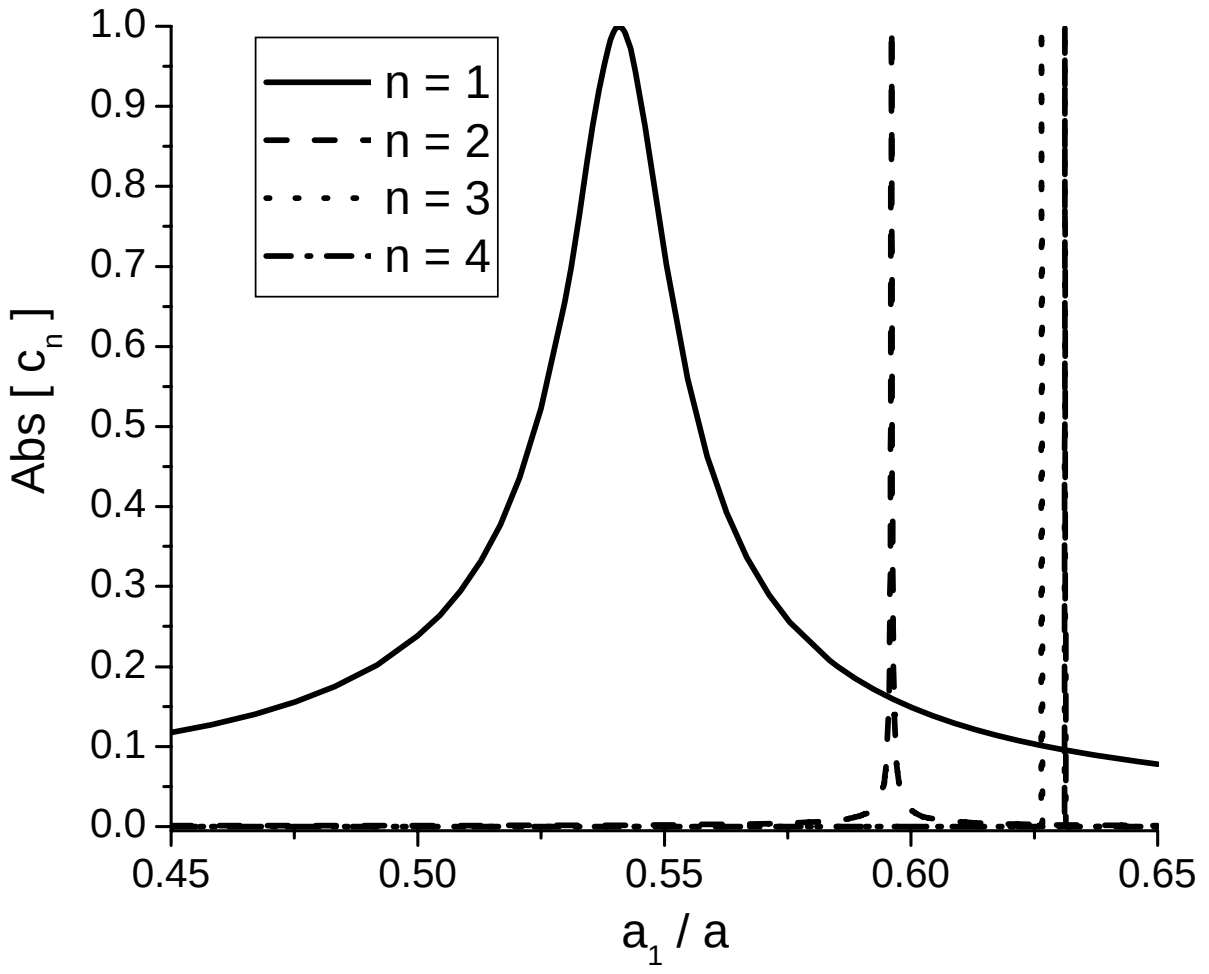

Figure 7

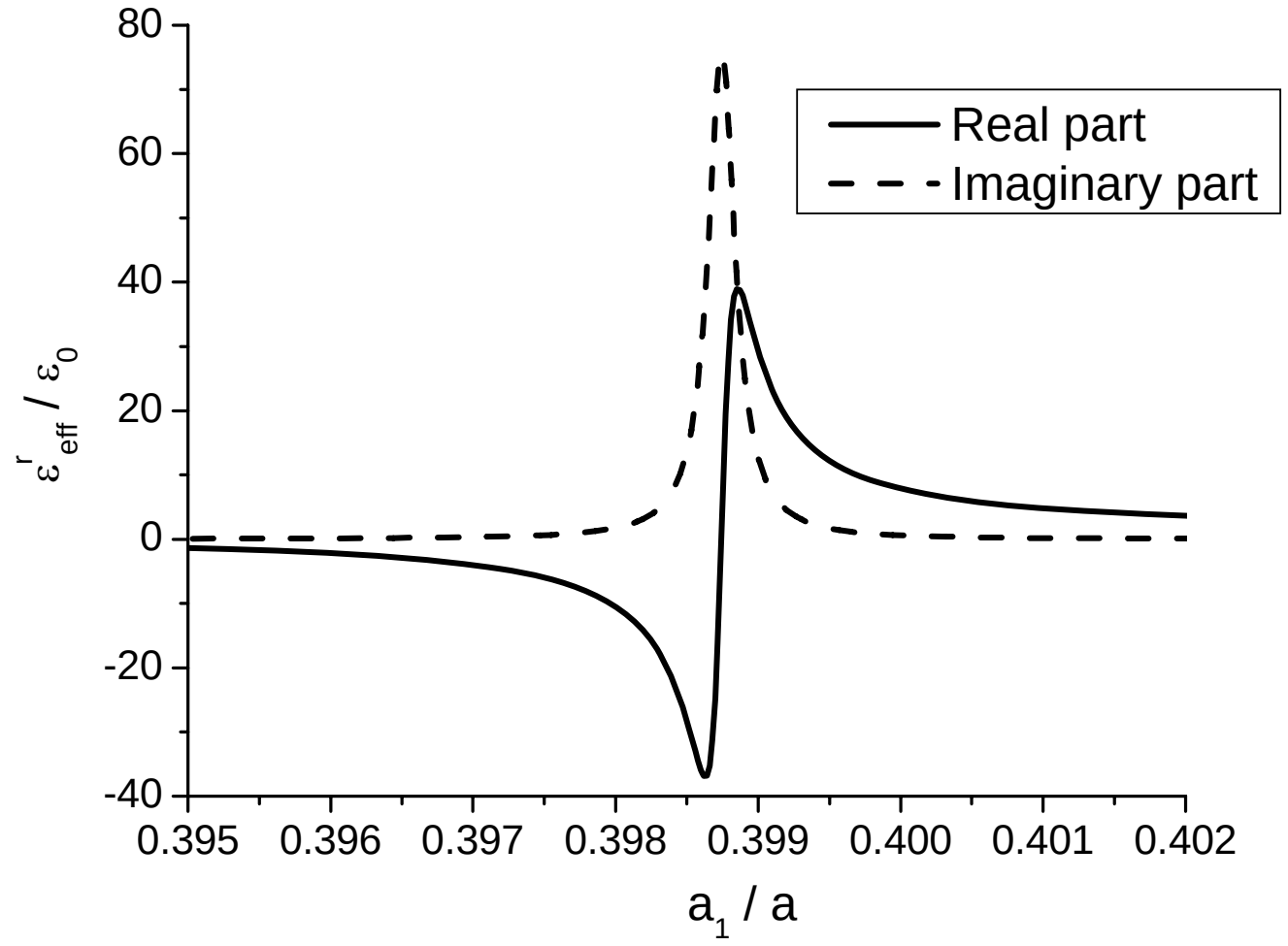

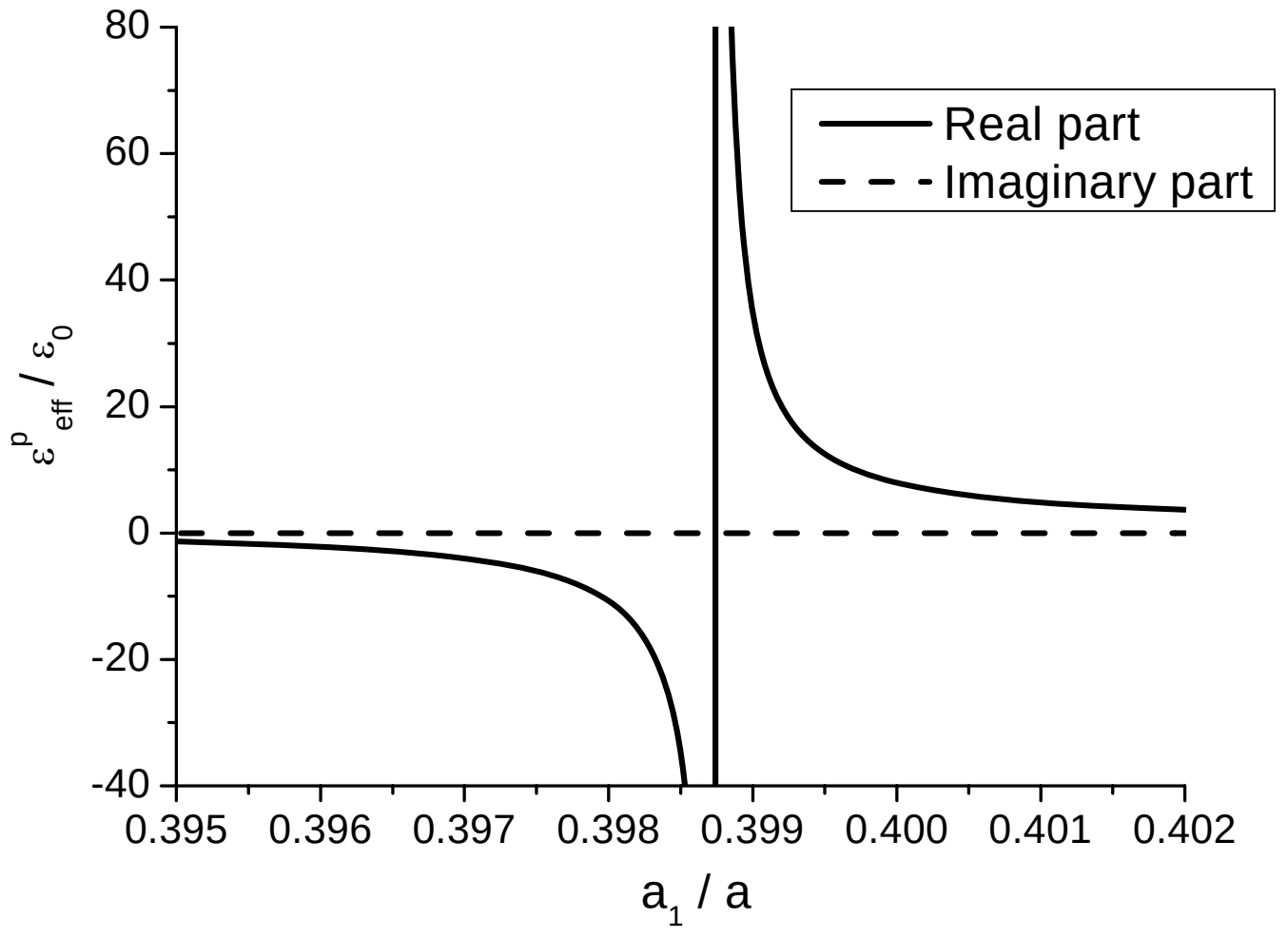

Figure 8

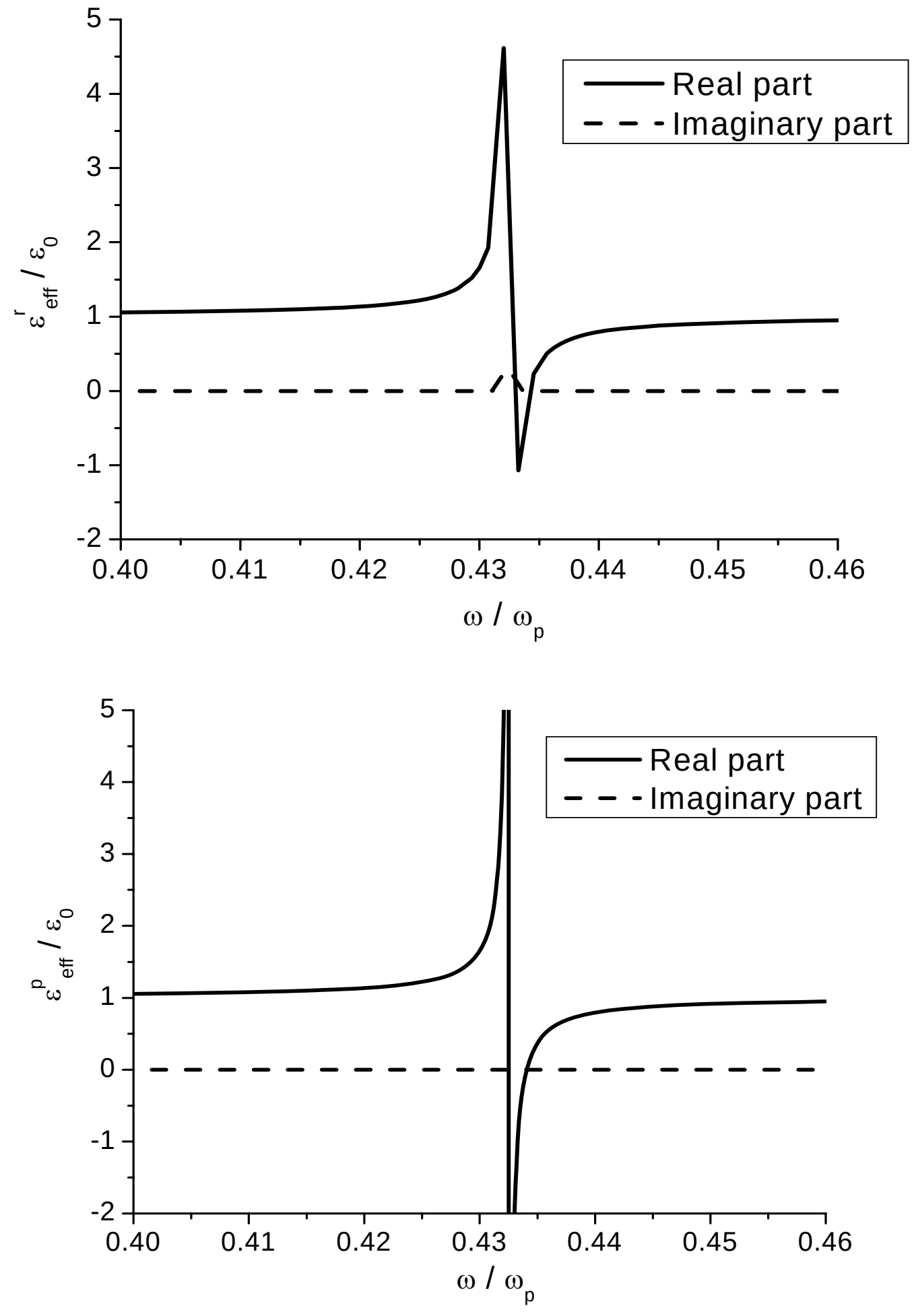

Figure 9